\documentstyle[preprint,aps,tighten,epsfig,amsmath,floats]{revtex}
\newcommand{\ds}{\displaystyle}
\newcommand{\del}{\mbox{\boldmath $\nabla$}}
\newcommand{\KL}{K^\Lambda}
\newcommand{\ML}{M^\Lambda}
\newcommand{\KLi}{K_i^\Lambda}
\newcommand{\Mi}{M_i}
\newcommand{\bKLi}{\bar{K}_i^\Lambda}
\newcommand{\tKLi}{\tilde{K}_i^\Lambda}

\newcommand{\tK}{\tilde{K}}
\newcommand{\bKL}{\bar{K}^\Lambda}
\newcommand{\tKL}{\tilde{K}^\Lambda}
\newcommand{\tM}{\tilde{M}}
\newcommand{\tML}{\tilde{M}^\Lambda}
\newcommand{\p}{{\bf p}}

\newcommand{\ben}{\begin{displaymath}}
\newcommand{\een}{\end{displaymath}}
\newcommand{\be}{\begin{equation}}
\newcommand{\ee}{\end{equation}}
\newcommand{\bea}{\begin{eqnarray}}
\newcommand{\eea}{\end{eqnarray}}
\newcommand{\eqn}[1]{\label{#1}}
\newcommand{\eq}[1]{Eq.~(\ref{#1})}

\newcommand{\fign}[1]{\label{#1}}
\newcommand{\fig}[1]{Fig.~\ref{#1}}

\begin{document}
\draft \title{Three-body system in leading order Effective Field Theory without
three-body forces}
\author{B. Blankleider and J. Gegelia}
\address{Department of Physics, The Flinders University of South Australia, \\
Bedford Park, SA 5042, Australia}
\date{\today}
\maketitle 

\begin{abstract}

The use of leading order effective field theory (EFT) to describe
neutron-deuteron scattering leads to integral equations that have unusual
behaviour: when only two-body interactions are included, the scattering
amplitude does not approach a limit when the cutoff used to solve the equations
is removed. It has recently been shown that this cutoff dependence can be
eliminated by the careful inclusion of a three-body force. In this paper we show
that the cutoff dependence is just a reflection of the fact that the
aforementioned integral equations admit an infinite number of solutions amongst
which only one corresponds to the physical scattering amplitude.  We show how to
numerically extract the physical scattering amplitude from the general solution
and in this way explicitly demonstrate that the amplitude for a particle
scattering off a two-body bound state, in leading order EFT, is in fact
determined entirely by two-body forces.

\end{abstract}

\pacs{
03.65.Nk,  
%Nonrelativistic scattering theory
11.10.Gh,  
%Renormalization
12.39.Fe,  
%Chiral Lagrangians
13.75.Cs.} 
%Nucleon-nucleon interactions (including antinucleons, deuterons,
%          etc.)

\section{Introduction}

Effective field theory (EFT) approaches to problems in nuclear physics have been
under intensive investigation during the last few years. A review of recent
developments (and references to the relevant papers) can be found in
\cite{bira}. In the two-body sector, most of the theoretical problems
encountered initially have now been resolved with recent EFT calculations of
nucleon-nucleon scattering giving a particularly successful description
\cite{Ordonez,Epelbaum,Gegelia:1998xr,Gegelia:1999ja,Park}.
On the other hand, application of the EFT program to the three-body sector
remains problematic. Here we would like to address an especially intriguing
problem encountered in the lowest order EFT calculation of neutron-deuteron
scattering in the $J=1/2$ channel \cite{BHK_PRL}. 

For very low energy EFT calculations of nuclear systems, one integrates out all
particles other than the nucleon, and at leading order one is left with just a
constant contact interaction between the nucleons. A problem arises immediately
when one tries to use this contact interaction in a standard calculation of
three-body scattering. In this regard we note that the spinor nature of the
nucleons plays no essential role here and that exactly the same problem occurs
in the case where the three particles are all scalars. Thus for the sake of
simplicity and without loosing generality, we shall present our discussion for
the case of three bosons, described by EFT, where at very low energies all other
particles have been integrated out and the leading order interaction is a
constant contact term.

The problem in question can be described as follows. To find the leading order
amplitude for a boson scattering off a two-boson bound state we need to sum up
an infinite number of diagrams involving the two-body constant contact
interaction.  While each leading order three-body diagram with re-summed
two-body interactions is individually finite, when one tries to sum up these
diagrams by using an integral equation
\begin{equation}
a(p,k) = M(p,k)+\int_0^\infty dq\ Z(p,q) a(q,k), \eqn{a}
\end{equation}
where $a$ is the summed amplitude, $M$ is the Born term, and $Z$ is the kernel,
one finds that the operator $(1-Z)^{-1}$ does not exist. Thus one cannot express
the amplitude as $a= (1-Z)^{-1}M$ with the consequence that the numerical
solution of \eq{a} is particularly difficult to construct.

If one tries to handle this problem by introducing an ultraviolet cutoff for the
integral in \eq{a}, one then finds that although the modified equation is easily
solved (e.g.\ by matrix inversion), its solution is sensitive to the chosen
cutoff and the limit as the cutoff is removed does not exist.  To resolve this
problem, Bedaque {\em et al.}\ \cite{BHK_PRL} introduced a one-parameter
three-body force counter-term into their leading order EFT calculations. Indeed,
they argued that the introduction of this three-body force is a necessary and
sufficient condition to eliminate the cut-off dependence.

%The present paper studies the problems of $J=1/2$ neutron-deuteron scattering
%from effective field theory at very low energies. 
%The $J=1/2$ neutron-deuteron scattering amplitude of EFT with only nucleons as
%explicit degrees of freedom satisfies the equations of
%Skornyakov and Ter-Martirosyan. The analogous equation with analogous
%properties 
%emerges for three-boson
%system.  

In the present paper we solve the problem of sensitivity to the ultraviolet
cut-off without introducing three-body forces into the leading order Lagrangian.
We do this by recognising that \eq{a} has an infinite number of solutions, only
one of which corresponds to the physical amplitude for particle-bound state
scattering. Furthermore, we show that the introduction of a particular
ultraviolet cutoff into \eq{a} results in an equation whose solution is an
approximation to just {\em one} of the infinite number of solutions of the
original \eq{a}. Moreover, by varying the ultraviolet cutoff, we end up
obtaining approximations that jump between {\em different} solutions of \eq{a}.
It is this jumping between solutions which is responsible for the sensitivity to
the ultraviolet cut-off observed in Ref.\ \cite{BHK_PRL}. By performing a
straightforward numerical analysis of the solutions with different cutoff
parameters, we are able to construct the actual physical solution of \eq{a}
where the input consists of two-body interactions only. In this way the present
paper extends the ideas presented in Ref. \cite{jambul} and also complements
these through explicit numerical calculations.

Although we have considered the case of three boson scattering for simplicity,
exactly the same considerations hold for the problem of $J=1/2$ channel
neutron-deuteron scattering in the framework of EFT. We will address the
specific case of neutron-deuteron scattering in a separate paper.

\section{Three-body system in leading order EFT}
\subsection{Integral equations for boson-dimeron scattering}

The leading order Lagrangian for the considered EFT of non-relativistic
self-interacting bosons is given by  \cite{BHK_NP}
\begin{equation}
\label{lag}
{\cal L}  =  \phi^\dagger
            \left( i\partial_{0}+\frac{\del^{2}}{2m}\right)\phi
 - \frac{C_0}{2} (\phi^\dagger \phi)^2
% - \frac{D_0}{6} (\phi^\dagger\phi)^3,\nonumber
\end{equation}
where $\phi$ is the boson field, $m$ is its mass, and $C_0$ is a coupling
constant.  For the sake of convenience \cite{Kaplan} one can rewrite this theory
by introducing a dummy field $\Phi$ with the quantum numbers of two bosons,
referred to as a ``dimeron'' \cite{BHK_NP}:
\begin{equation}
\eqn{lagt}
{\cal L}=\phi^\dagger
             \left( i\partial_{0}+\frac{\del^{2}}{2m}\right)\phi
         + \Delta \Phi^\dagger \Phi
  -\frac{g}{\sqrt{2}} (\Phi^\dagger \phi\phi +\mbox{h.c.}).
\end{equation}
\noindent
The scale parameter $\Delta$ is included to give the field $\Phi$ the usual mass
dimension of a heavy field.  Observables depend on the parameters of
\eq{lagt} only through the combination $C_0\equiv g^2/\Delta$.

The (bare) dimeron propagator is a constant $i/\Delta$ and the boson propagator
is given by the usual non-relativistic expression $i/(p^0-p^2/2m+i\epsilon)$
where $p=|\p|$ (similar three-vector notation is used below for other momentum
variables).  The dressing of the dimeron propagator is illustrated in
\fig{fig1}(a). Summing loop-diagrams, subtracting the divergent integral at
$P^0=P=0$ (where $P^0$ and $P=|{\bf P}|$ refer to the momentum of the dimeron)
and removing the cut-off, one gets the following dressed dimeron propagator
\cite{BHK_NP}:
\begin{equation}
\label{Dprop}
i S(P^0,P) =  \frac {-i}{\ds - \Delta^R + \frac{m g^{2}}{4\pi}
               \sqrt{-m P^0+P^2/4-i\epsilon} +i\epsilon} 
\end{equation}
where $\Delta^R$ is the renormalised parameter ($\Delta$ has absorbed the linear
divergence).  Attaching four boson lines to this dressed dimeron propagator one
gets the two-particle scattering amplitude at leading order. This amplitude
has the form of an effective range expansion truncated at leading order with
$g^2/\Delta^R = 4\pi a_2/m$ where $a_2$ is the two-body scattering length.

\begin{figure}[t]
\centerline{\epsfxsize=14cm\epsfbox{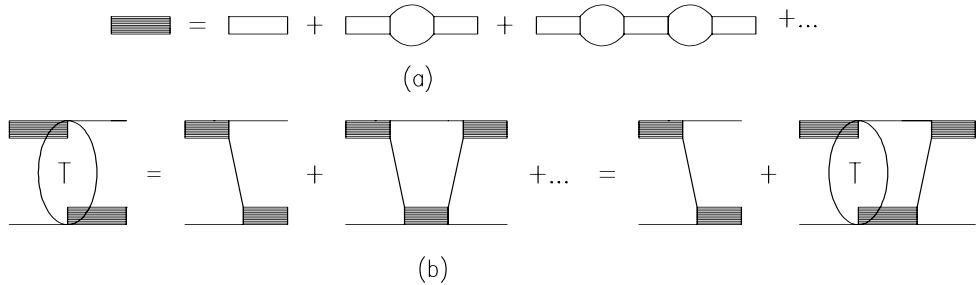}}
\vspace*{0.5cm}
\caption{(a) Dressing of the dimeron. (b) Diagrams contributing to
particle - dimeron scattering.}
\fign{fig1}
\end{figure}
For the scattering of a particle off a two-body bound state, standard
power counting shows that the leading order contribution to the amplitude $T$ is
given by the diagrams illustrated by the first equality of \fig{fig1}(b). The
sum of all these diagrams satisfies the equation represented by the second
equality in \fig{fig1}(b). For $s$-wave scattering it is convenient to
define the function $a(p,k)$ in terms of the $s$-wave amplitude $T_0(p,k)$ by
\begin{equation}
a(p,k)=\frac{p^2-k^2}{-1/a_2+\sqrt{{3p^2/4}-mE}}\,\frac{T_0(p,k)}{mg^2}
\label{defa}
\end{equation}
which can be shown to satisfy the equation
\cite{ST,Bedaque,BK,BHK_PRC}
\begin{equation}
a(p,k) = M(p,k)+\frac{2\lambda}{\pi}\int_0^\infty dq\ M(p,q)
\frac{q^2}{q^2-k^2-i\epsilon} a(q,k),         \eqn{aeq}
\end{equation}
where $k$ ($p$) is the incoming (outgoing) momentum magnitude, $E = 3k^2/4m -
1/ma_2^2$ is the total energy, and
\begin{equation}
M(p,q)= \frac{4}{3}\left(\frac{1}{a_2}+\sqrt{\frac{3}{4}p^2-m E}\,\,\right)
   \frac{1}{pq}{\rm ln}
    \left(\frac{q^2+p q+p^2-m E}
               {q^2-q p+p^2-m E}\right).
%\equiv \frac{8}{3}\left(\frac{1}{a_2}+\gamma_k(p)\right)L_E(p,q).
\end{equation}
%where the second equality introduces short notations for long expressions.
\eq{aeq} was first derived by Skorniakov and Ter-Martirosian \cite{ST}
(S-TM equation) and has $\lambda=1$ for the three-boson case.  Three nucleons in
the spin $J=1/2$ channel obey a pair of integral equations with similar
properties to this bosonic equation, and the spin $J=3/2$ channel corresponds to
$\lambda=-1/2$.

It was shown by Danilov \cite{danilov} that for $\lambda =1$ the homogeneous
equation corresponding to \eq{aeq},
\begin{equation}
a_h(p,k) = \frac{2\lambda}{\pi}\int_0^\infty dq\ M(p,q)
\frac{q^2}{q^2-k^2-i\epsilon} a_h(q,k),         \eqn{aheq}
\end{equation}
has a solution for arbitrary $E$. In particular, there exists a solution of the
homogeneous equation for every energy corresponding to the {\em scattering} of a
projectile off a two-body bound state. Although all such scattering energy
solutions are unphysical, it must be emphasised that they are purely an artifact
of having two-body effective potentials that are zero-range ($\delta$-function
potentials in coordinate space). By contrast, the solution of the corresponding
homogeneous equation for non-zero range potentials (namely the bound state
Faddeev equation) has no solutions for energies corresponding to scattering. In
turn, it should be remembered that zero range effective potentials are
themselves an artifact of restricting the EFT model to the lowest order terms -
they are not a property of the full EFT approach \cite{Gegelia:1998xr}.

Although unphysical, the existence of a scattering energy solution $a_h$ of
\eq{aheq} has practical consequences for finding the physical scattering
amplitude $a_{ph}$ satisfying \eq{aeq}. The problem is that the existence of
$a_h$ implies that \eq{aeq} has an infinite number of solutions given by $a =
a_{ph} + C a_h$ where $C$ is an arbitrary parameter. As we shall see, \eq{aheq}
has actually more than one solution for any given $E$. Writing these solutions
as $a_h^i$ where $i=1,2,3,\ldots$, the most general solution of \eq{aeq} can
therefore be written as
\begin{equation}
a = a_{ph} + \sum_i C_i a_h^i \ . \eqn{a_general}
\end{equation}
Thus the sum of the diagrams in \fig{fig1}(b), which {\em defines} the physical
amplitude $a_{ph}$, is only one of an infinite number of solutions to \eq{aeq}.

Setting $\lambda=1$ and writing \eq{aeq} and \eq{aheq} in operator form as
\begin{equation}
a = M + MG_0 a   \eqn{aop}
\end{equation}
and
\begin{equation}
\quad a_h=MG_0 a_h  \eqn{ahop}
\end{equation}
respectively, it is clear from the existence of a non-zero solution $a_h$ of
\eq{ahop} that the inverse operator $(1-MG_0)^{-1}$ does not exist. This in turn
means that one cannot write the solution of \eq{aop} as $a=(1-MG_0)^{-1}M$ with
the practical consequence that \eq{aeq} cannot be solved by matrix inversion.
Indeed we find that many other numerical methods, for example Pad\'{e}
approximants, are likewise unstable for this case.

The task of finding the physical amplitude $a_{ph}$ appears to be formidable.
Even if \eq{aeq} could be solved, one would still need to determine the
appropriate values of the parameters $C_i$ in order to extract $a_{ph}$.
Attempts to find $a_{ph}$ by writing it directly as the sum
\begin{equation}
a_{ph}= M+MG_0M+MG_0MG_0M+\ldots
\end{equation}
also do not help, as this sum cannot be evaluated numerically due to extreme
sensitivity to roundoff errors. Again, this numerical instability appears to be
linked to the non-existence of $(1-MG_0)^{-1}$.  Fortunately, all these
numerical difficulties can be overcome. The rest of this paper is devoted to
accomplishing this task.

\subsection{Extracting the physical boson-dimeron amplitude}
\subsubsection{Asymptotic behaviour}

In order to distinguish the physical boson-dimeron amplitude $a_{ph}$ from the
infinite number of non-physical solutions given by \eq{a_general}, it is useful
to examine the asymptotic behaviour of the general solution $a(p,k)$ to the
S-TM equation for large $p$. It has the form \cite{danilov}
\begin{equation}
a(p,k)= \sum_i A_i\left( k\right)p^{s_i}+O\left( {1/p}\right)
\eqn{aasymptotic}
\end{equation}
where $s_i$ are roots of the equation
\begin{equation}
\label{seq}
1- \frac{8\lambda}{\sqrt{3}}
   \frac{\sin\pi s/6}{s \cos\pi s/2}=0.
\end{equation}
The summation in \eq{aasymptotic} goes over all solutions of
Eq. (\ref{seq}) for which $|{\rm Re} s|<1$.
For $\lambda =1$ \eq{seq} has two roots for which $|{\rm Re} s|<1$:
$s=\pm is_0$,
where $s_0\approx 1.00624$, so that  \eq{aasymptotic} gives the
asymptotic behaviour of the boson-dimeron amplitude as
\begin{equation}
a(p,k)\sim A_1\left( k\right)p^{is_0}+A_2\left( k\right)p^{-is_0}
\label{aasymptotics}.
\end{equation}
%One of the arbitrary constants $A_1(k)$ and $A_2(k)$
%is determined by the other when this solution is joined to the solution in the
%region of small $p$. Hence the solution of \eq{aeq} depends on a single
%arbitrary parameter. 
It can be shown \cite{danilov} that for $\lambda=1$
the asymptotic behaviour of the solution $a_h(p,k)$ of the homogeneous equation,
\eq{aheq}, is also given by the right hand side of \eq{aasymptotics}.

What distinguishes the physical amplitude $a_{ph}$ from any other amplitude $a$
satisfying the S-TM equation is that it is the sum of the diagrams in
\fig{fig1}(b); that is, it is equal to the sum of the series obtained by
iterating \eq{aeq}. In this way $a_{ph}$ can be expressed as a power series in
the parameter $\lambda$. By contrast, it can easily be seen that for the
physical amplitude $a_{ph}$, the right hand side of \eq{aasymptotics} must be
identically zero, i.e., it cannot be expressed as a non-trivial power series in
$\lambda$.\footnote{For if $A_1$ and $A_2$ were non-vanishing and we could write
$s_0(\lambda) = c_0 + c_1\lambda+c_2\lambda^2+\ldots$ , then substituting this
into \eq{aasymptotics} we would obtain a power series in $\lambda$ whose
coefficients do not have the same asymptotic $p$ behaviour as the corresponding
diagrams coming from the iteration of \eq{aeq}. } Thus
\begin{equation}
a_{ph}(p,k) \to 0\quad \mbox{as}\quad p\to \infty.
\end{equation}

\subsubsection{Unitarity}

Another essential property of the physical amplitude $a_{ph}$ is that it
satisfies unitarity; that is, the on-shell physical amplitude $a=a_{ph}$
satisfies the relation 
\begin{equation}
a(k,k) - a^*(k,k)=  2i\lambda k\left|a(k,k)\right|^2.
\eqn{unitarity}
\end{equation}
In this respect it is important to note that if an amplitude $a$ satisfies the
S-TM equation, it does {\em not} necessarily mean that $a$ satisfies
unitarity. Indeed, if we write the S-TM equation in operator form as $a=M+MG_0a$
and then try to prove unitarity in the usual way, we would firstly want to write
$M^{-1}=a^{-1}+G_0$, then subtract the Hermitian conjugate of this equation, and
lastly, rearrange to obtain the unitarity relation $a-a^\dagger =
a(G_0-G_0^\dagger)a^\dagger$ which reduces to \eq{unitarity} on-shell. However,
if the inverse $(1-MG_0)^{-1}$ doesn't exist, then it is easy to show that
either operator $a$ has no right-inverse or $M$ has not left-inverse. In either
case, the usual proof of unitarity breaks down at the first step.  On the other
hand, because $a_{ph}$ can be expressed as an iteration of the S-TM equation,
$a_{ph}=M+MG_0M + MG_0MG_0M +\ldots$, the unitarity relation for $a_{ph}$ can
easily be proved directly without the need to take the inverse of any operator.

As the physical amplitude satisfies unitarity, we would like to consider only
those solutions $a$ of the S-TM equations that likewise satisfy unitarity.  For
\eq{a_general}, this means a strong restriction on the allowed values for the
parameters $C_i$. This restriction on the solutions of \eq{aeq} can be achieved
through the introduction of the $K$ matrix equation
\begin{equation}
K(p,k) = M(p,k)+\frac{2\lambda}{\pi}P\int_0^\infty dq\ M(p,q)
\frac{q^2}{q^2-k^2} K(q,k),         \eqn{Keq}
\end{equation}
where $P$ stands for ``principal value'' and where only the real solutions for
the $K$ matrix $K(p,k)$ are considered.  It is then easy to show that the
amplitude
\begin{equation}
a(p,k)\equiv \frac{K(p,k)}{1-i\lambda kK(k,k)}
\label{amplk}
\end{equation}
satisfies both the S-TM equation and the unitarity relation of \eq{unitarity}.
Conversely, any amplitude $a(p,k)$ that satisfies the S-TM equation and is
unitary can be shown to be of the form given by \eq{amplk} where $K$ is real and
satisfies \eq{Keq}.

One might hope that the unitarity condition would single out the physical
solution from an infinite number of solutions of \eq{aeq}.  This would require
\eq{Keq} to have a unique solution corresponding to the physical amplitude.
Unfortunately the homogeneous equation corresponding to \eq{Keq} admits
non-trivial solutions, and this means that \eq{Keq} has in fact an infinite
number of solutions.  Nevertheless, we know from the above discussion of the
asymptotic behaviour of $a_{ph}$ and the relation between the $a$ amplitude and
the $K$ matrix, \eq{amplk}, that amongst all the solutions to \eq{Keq}, the one
that corresponds to the physical amplitude is the solution which vanishes for
large $p$.

\section{Numerical solution}
\subsection{Solution to the inhomogeneous equation}

As we are specifically interested in the description of three bosons within
leading order EFT, we shall implicitly assume that $\lambda=1$ in all the
equations below. Writing \eq{Keq} symbolically for this case as $K=M+MG_0^PK$,
the fact that the homogeneous equation corresponding to this equation has a
solution for an arbitrary value of the energy means that the inverse operator
$(1-MG_0^P)^{-1}$ does not exist.  Thus trying to solve \eq{Keq} directly
presents exactly the same type of numerical difficulties as discussed above for
the case of \eq{aeq}.  In order to solve \eq{Keq} numerically we introduce
regularisation such that, in contrast to \eq{Keq}, the regularised equation has
a unique solution.  Although this solution depends strongly on the cutoff
parameter(s) of the regularisation, we will see that this is {\em not} an
indication of any difficulty with leading order EFT, as suggested in Ref.\
\cite{BHK_PRL}, but rather it is simply a consequence of the fact that different
regularisations correspond to different solutions of the unregularised equation.

We consider the regularised equation
\begin{equation}
\KL(p,k)
=M(p,k)+\frac{2\lambda}{\pi} \ P \int_0^\infty dq\ M^\Lambda (p,q)
\frac{q^2}{q^2-k^2} \KL(q,k)  \eqn{KLeq}
\end{equation}
where
\begin{equation}
\label{regkernelaeq}
M^\Lambda (p,q)=M(p,q)+\frac{1}{\Lambda+q^2}.
\end{equation}
For this regularisation the inverse operator $(1-M^\Lambda G_0^P)^{-1}$ exists
and \eq{KLeq} is easily solved.  We find that the solution of this equation,
$K\equiv\KL$, satisfies \eq{Keq} with very good accuracy for any $\Lambda$ that
is sufficiently large; moreover, by increasing $\Lambda$ one can obtain
solutions $\KL$ that will satisfy \eq{Keq} to any given accuracy. As expected
from the asymptotic behaviour specified by \eq{aasymptotics}, we find that
$\KL(p,k)$ has oscillating behaviour for large $p$ - see \fig{KLfig}. We also
find that $\KL (p,k)$ has an oscillating behaviour with respect to $\Lambda $,
in agreement with what was observed in Ref.\ \cite{BHK_PRL}.
\begin{figure}[t]
\centerline{\epsfxsize=8cm\epsfbox{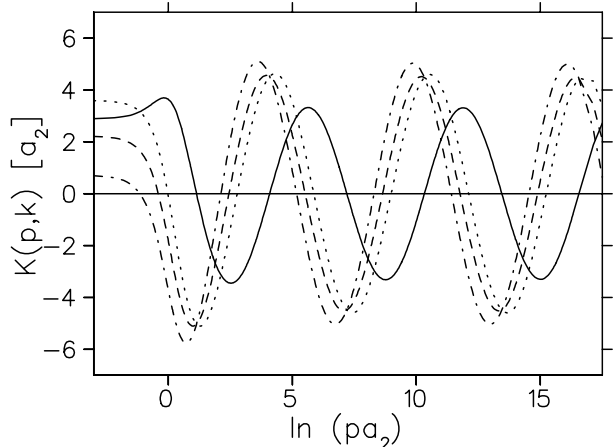}}
\caption{The dot-dashed, dashed, and dotted lines are solutions
$K^{\Lambda_i}(p,k)$ of the inhomogeneous $K$ matrix equation, \eq{KLeq},
corresponding to three different cutoffs $\Lambda_1= 1.6\times 10^7 a_2^{-1}$,
$\Lambda_2=3.2\times 10^7a_2^{-1}$ and $\Lambda_3=6.4\times 10^7a_2^{-1}$,
respectively.  The solid line is the function
$K_h(p,k)=K^{\Lambda_1}(p,k)-K^{\Lambda_2}(p,k)$ which forms a solution of the
corresponding homogeneous equation. In all cases $k=0.82a_2^{-1}$.}
\fign{KLfig}
\end{figure}

More information on the functional form of $\KL(p,k)$ can be obtained by
constructing solutions $K^{\Lambda_i}(p,k)$ ($i=1,\ldots,4$) corresponding to
four different values of $\Lambda$.  As each of the $K^{\Lambda_i}(p,k)$, for
$\Lambda_i$ large enough, is a solution to \eq{Keq}, the differences
$K^{\Lambda_i}(p,k)-K^{\Lambda_j}(p,k)$ are clearly solutions to the homogeneous
equation corresponding to \eq{Keq}. However, what is particularly interesting
about such differences is that their ratio
$$
\frac{K^{\Lambda_1} (p,k)-K^{\Lambda_2} (p,k)}
{K^{\Lambda_3} (p,k)-K^{\Lambda_4} (p,k)}
$$ 
is found to be totally independent of the momentum variable $p$. From this
observation we conclude that the structure of $\KL$ must be of the form
\begin{equation}
\KL(p,k)=K_0(p,k)+C(\Lambda)K_h(p,k)
\label{Kstruc}
\end{equation}
where $K_0(p,k)$ is a solution to \eq{Keq}, $K_h(p,k)$ is a solution to the
corresponding homogeneous equation, and $C(\Lambda)$ is purely a function of
$\Lambda$ (we consider the initial momentum $k$ to be fixed). As the
normalisation of the function $K_h(p,k)$ is not determined by the homogeneous
equation, for definiteness we take
$K_h(p,k)=K^{\Lambda_1}(p,q)-K^{\Lambda_2}(p,q)$ which is drawn in \fig{KLfig}
together with three of the particular solutions to the inhomogeneous equation.
Note that $K_0(p,k)$ and $K_h(p,k)$ do not depend on $\Lambda$. Also, because
each $\KL(p,k)$ oscillates as a function of $p$ with a phase that depends on
$\Lambda$, $K_0(p,k)$ and $K_h(p,k)$ must similarly oscillate but with differing
phases.  Being a solution to the inhomogeneous $K$ matrix equation, $K_0(p,k)$
can be written generally as $K_{ph}(p,k)+K_h^*(p,k)$ where $K_{ph}$ is the
physical $K$ matrix which has the vanishing asymptotical behaviour (i.e. the $K$
matrix corresponding to the physical amplitude $a_{ph}$), and $K_h^*(p,k)$ is
another solution to the homogeneous equation. We may thus write the solution to
\eq{Keq} for large $\Lambda$ as $K\equiv \KL$ where
\begin{equation}
\KL(p,k)=K_{ph}(p,k)+K_h^*(p,k)+C(\Lambda )K_h(p,k).
\label{KKstruc} 
\end{equation}
For large $p$ the structure of \eq{KKstruc} agrees with the asymptotical
behaviour of the solution to the regularised equation obtained in
\cite{BHK_NP}:\footnote{Note that the cutoff used in Ref.\
\cite{BHK_NP} is different from ours; however, the functional form of the
asymptotical behaviour does not depend on the particular choice of the cutoff.}
\begin{equation}
\KL(p\gg 1/a_2, k)=-\frac{\gamma}{\cos\left[ s_0\ln\left( p_*a_2\right)+
\epsilon \right]}\cos\left( s_0\ln\frac{p}{p_*}\right)
\label{regasym}
\end{equation}
where $p_*=\exp\left( -\delta/s_0\right)\Lambda$
and $\delta$, $\gamma $ and $\epsilon $ are cutoff-independent constants. 
Indeed, by writing \eq{regasym} as
\begin{equation}
\KL(p\gg 1/a_2, k)=-\gamma \cos\left[ s_0\ln \left( pa_2\right)
+\epsilon\right]-\gamma \tan\left[s_0\ln\left( \Lambda a_2\right)-\delta
+\epsilon\right]\sin\left[s_0\ln \left(pa_2\right)+\epsilon\right]
\label{regasymrewritten}
\end{equation} 
and noting that the physical $K$ matrix $K_{ph}$ vanishes for large $p$, we can
deduce that $K_h^*(p,k)\sim \cos\left[s_0\ln \left( pa_2\right)
+\epsilon\right]$, $K_h(p,k)\sim \sin\left[s_0\ln \left( pa_2\right)
+\epsilon\right]$ and $C(\Lambda )\sim \tan \left[s_0\ln\left( \Lambda
a_2\right)-\delta +\epsilon\right]$. The deduced asymptotic behaviour of $K_h$
and $K_h^*$ is consistent with \eq{aasymptotics} and is further borne out by our
numerical results.

% ak gadis sazgvari!!!!!!!!!!!!!!!!!!!!!!!!!!!!!!!!!!!!!!!!!!!!!!!!!!!!!!!

The existence of (at least) two linearly independent solutions to the
homogeneous equation may seem a little surprising in light of the fact that
\eq{Kstruc} contains just one solution, $K_h$, with a $\Lambda$-dependent
coefficient. To further check that the structure given by \eq{Kstruc} is 
consistent with there being more than one linearly independent solution
to the homogeneous equation, we consider the equation  
\begin{equation}
\label{tKeq}
\tK(p,k)
=M(p,k)+\frac{2\lambda}{\pi} \ P \int_0^\infty dq\ \tM(p,q)\tK(q,k),
\end{equation}
where
\begin{equation}
\label{araertgv}
\tM(p,k)= \frac{2}{\sqrt{3}}
   \frac{1}{q}{\rm ln}
    \left(\frac{q^2+p q+p^2}{q^2-p q+p^2}\right).
\end{equation}
\noindent
To obtain $\tM$ we substituted $a_2=\infty$ and $mE=0$ in the expression for
$M(p,q)$, and also removed $q^2/(q^2-k^2)$ from the integrand.  The homogeneous
equation corresponding to \eq{tKeq} has solutions and hence \eq{tKeq} cannot be
solved directly. We again introduce regularisation which enables a
straightforward numerical
solution. Thus we solve the equation
\begin{equation}
\eqn{tKLeq}
\tKL(p,k)
=M(p,k)+\frac{2\lambda}{\pi} \ P  \int_0^\infty dq\ \tML  (p,q)
\tKL (q,k),
\end{equation}
where
\begin{equation}
\label{regkernel}
\tML(p,q)=\tM(p,q)+\frac{1}{\Lambda+q^2}.
\end{equation}
\noindent
For a sufficiently large cutoff $\Lambda$, the solution
$\tK(p,k)\equiv\tKL(p,k)$ satisfies \eq{tKeq} with very good accuracy and has
oscillating behaviour for large $p$. $\tKL(p,k)$ is also oscillating with
respect to $\Lambda $. Just like the $\KL$ of the original problem, the
solutions $\tKL$ have the property that the ratio
$$
\frac{\tK^{\Lambda_1}(p,k)-\tK^{\Lambda_2}(p,k)}
{\tK^{\Lambda_3}(p,k)-\tK^{\Lambda_4}(p,k)}
$$ 
does not depend on the momentum $p$. The amplitude structure is given
accordingly by
\begin{equation}
\tKL(p,k)=\tK_0(p,k)+A(\Lambda )\tK_h(p,k)
\label{tKstruc}
\end{equation}
where $\tK_0(p,k)$ is a solution to \eq{tKeq} and $\tK_h(p,k)=
\tK^{\Lambda_1}(p,k)-\tK^{\Lambda_2}(p,k)$ is a solution to the corresponding
homogeneous equation. Everything is the same as before except that now one can
actually solve the homogeneous equation analytically. We find that there are
{\em two} linearly independent solutions, $\sin\left[s_0\ln(a_2p)\right]$ and
$\cos\left[s_0\ln(a_2p)\right]$, even though only one solution enters with a
$\Lambda$-dependent coefficient in the general numerical form given by
\eq{tKstruc}. It is easy to see that the other linearly independent solution
contributes into $\tK_0(p,k)$ and is responsible for its oscillating behaviour.

\subsection{Extracting the physical amplitude}

In the previous subsection we have shown that the inhomogeneous $K$ matrix
equation, \eq{Keq}, can be solved numerically by introducing a sufficiently
large cutoff $\Lambda$, and that the resulting solution is of the form given by
\eq{KKstruc}.  Writing \eq{KKstruc} as
\begin{equation}
\KL(p,k)=K_{ph}(p,k)+\KL_h(p,k),
\label{KLstruc} 
\end{equation}
we are left with the task of extracting the physical $K$ matrix $K_{ph}(p,k)$
from the numerical values for $\KL(p,k)$. Because $K_{ph}(p,k)$ vanishes for
large $p$, the function $\KL_h(p,k)$ in \eq{KLstruc} is that solution to the
homogeneous equation corresponding to \eq{Keq} which has the same asymptotic
behaviour as $\KL(p,k)$; that is, $\KL_h(p,k)$ satisfies the two equations
%\bea
%\KL_h &=&M G_0^P \KL_h, \\[3mm]
%\KL_h &\underset{p\rightarrow\infty}{\sim}& \KL.
%\eea
\begin{eqnarray}
\KL_h(p,k) &=& \frac{2\lambda}{\pi} \ P \int_0^\infty dq\ M(p,q)
\frac{q^2}{q^2-k^2} \KL_h(q,k), \eqn{KLheq} \\[3mm]  
\KL_h(p,k) &\underset{p\rightarrow\infty}{\sim}& \KL(p,k). \eqn{KLhsim}
\end{eqnarray}
Note that $\KL_h(p,k)$ cannot be obtained by solving the homogeneous equation
corresponding to \eq{KLeq}; this homogeneous equation has no non-trivial
solutions since the inverse operator $(1-\ML G_0^P)^{-1}$ exists. On the other
hand, since \eq{KLheq} has solutions for every energy, its numerical solution
presents a significant technical problem. For example, if one simply discretises
this homogeneous equation and then tries to solve the resulting simultaneous
equations, one encounters severe numerical difficulties coming from the fact
that the eigenvalues of the operator $(1-M G_0^P)$ include not only zero, but
also values infinitely close to zero.

To solve \eq{KLheq}, and moreover, to obtain the solution with the asymptotic
behaviour required by \eq{KLhsim}, we solve a sequence of inhomogeneous
equations, like \eq{KLeq}, but with progressively smaller inhomogeneous terms.
That is, we solve the equations
\begin{equation}
\KLi(p,k) = \Mi(p,k)+\frac{2\lambda}{\pi}P\int_0^\infty dq\ \ML(p,q)
\frac{q^2}{q^2-k^2} \KLi(q,k),         \eqn{KLieq}
\end{equation}
where $i=1,2,3,\ldots$ and $\Mi(p,k)$ is taken to be a decreasing sequence
of functions. In particular, we take
\begin{equation}
\Mi(p,k)=\frac{a_2}{10^i(pa_2)^6+1}.    \eqn{Mi}
\end{equation}
Since $\Mi(p,k)$ becomes a vanishingly small function for increasing values of
$i$, we expect $\KLi(p,k)$, with $\Lambda$ large enough, to approximate
the solution of \eq{KLheq} for sufficiently large values of $i$.  At the same
time, since $\Mi(p,k)$ vanishes as $p\rightarrow \infty$, the solution
$\KLi(p,k)$ might be expected to have, up to some normalisation, the same
asymptotic behaviour (as a function of $p$) as $\KL(p,k)$.  That is in fact
what we find empirically, so that
\begin{equation}
\KLi(p,k) \underset{i\rightarrow \infty}{\sim} C_i \KL_h(p,k)
\end{equation}
where $C_i$ is a constant.

It is useful to illustrate the above procedure on the case of the homogeneous
equation corresponding to \eq{tKeq} for which an analytic solution is available.
Thus we would first like to consider the equation
\begin{equation}
\label{tKLieq}
\tKLi(p,k)
=\Mi(p,k)+\frac{2\lambda}{\pi} \ P \int_0^\infty dq\ \tML(p,q)
\tKLi (q,k),
\end{equation}
where $\Mi(p,k)$ is given by \eq{Mi} and $\Lambda$ has a large fixed
value. By the above discussion, we expect the solution to \eq{tKLieq} to
have the form
\begin{equation}
\tKLi(p,k)=A_i\tKL_h(p,k)+R^\Lambda_{i}(p,k)
\label{itersol}
\end{equation}
where $\tKL_h(p,k)$ is that solution of the homogeneous equation corresponding
to \eq{tKeq} which has the same asymptotic behaviour as the solution $\tKL(p,k)$
of \eq{tKLeq}, and $R^\Lambda_i (p,k)$ is a remainder term that becomes
progressively smaller with increasing $i$ and that decreases rapidly as a
function of $p$ beyond a close neighbourhood of the origin. The constant $A_i$
is expected to depend only on the choice of functions $\Mi(p,k)$. To see how
these expectations are borne out in practice, we have numerically solved
\eq{tKLieq} for the sequence of inhomogeneous functions given by \eq{Mi}.
We find that for different $i$ and up to a normalisation factor, the functions
$\tKLi(p,k)$ all have identical oscillating behaviour for large $p$. We scale
these functions so that their oscillating tails have the same amplitude, i.e.,
we rescale \eq{itersol} as
\begin{figure}[t]
\centerline{\epsfxsize=8cm\epsfbox{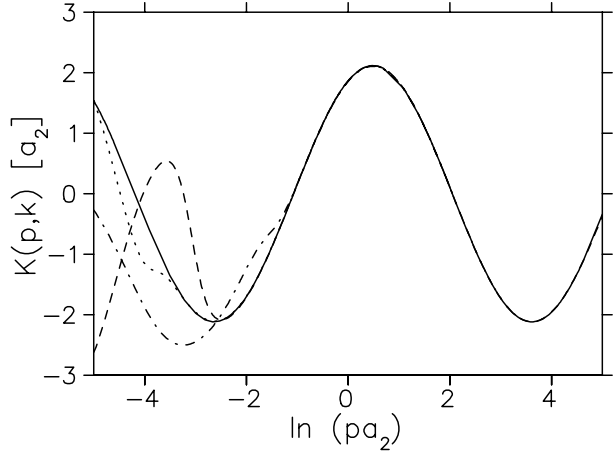}}
\vspace{2mm}

\caption{Solutions to the homogeneous equation corresponding to \eq{tKLeq} that
have the same asymptotic behaviour as the solution to \eq{tKLeq} with
$\Lambda=1.6\times 10^7 a_2^{-1}$.  The solid line is the exact solution
$\tKL_h$ to the homogeneous equation, the dot-dashed, dashed, and dotted lines
correspond to the sequential approximations $\bKL_i$ with $i=4$, $8$, and $10$,
respectively.}  \fign{figitersol}
\end{figure}
\begin{equation}
\bKLi(p,k)  = \tKL_h(p,k)+\bar{R}_i^\Lambda(p,k)
\eqn{itersolnorm}
\end{equation}
with $\bKLi\equiv\tKLi/A_i$ and $\bar{R}_i^\Lambda\equiv R_i^\Lambda/A_i$, and
then plot the results in \fig{figitersol}. We see that in agreement with our
expectations for increasing values of $i$, the remainder term
$\bar{R}_i^\Lambda$ becomes progressively smaller in such a way that two
sequential functions $\bKLi(p,k)$ and $\bKL_{i+1}(p,k)$ coincide beyond some
value of $p$ that is progressively approaching zero.  According to
\eq{itersolnorm} this tail, which is common for all functions $\bKLi$, is just
the tail of the solution to the homogeneous equation, $\tKL_h$. Choosing a
different sequence of inhomogeneous terms, for example
\begin{equation}
\Mi(p,k)=a_2\exp[-i(a_2p)^4],
\label{seqexp}
\end{equation}
we obtain a sequence of functions $\bKLi{'}$ that approaches exactly the same
solution $\tKL_h$ to the homogeneous equation. For this non-physical case we are
able to solve the homogeneous equation analytically. As the general solution
is a linear combination of the
independent
solutions $\sin[s_0\ln(a_2 p)]$ and $\cos[s_0\ln(a_2p)]$, we are  able to
write our sequence of solutions as
\begin{equation}
\tKLi(p,k) = A_i\sin[s_0\ln(a_2p) + \alpha] + R_i^\Lambda(p,k)
\label{itersolalt}.
\end{equation}
It is interesting to note that although the constant $\alpha$ depends on the
cutoff $\Lambda$, our numerical findings indicate that it does not depend on the
value of $i$. The function $\sin[s_0\ln(a_2p) + \alpha]$ (scaled appropriately)
is also plotted in \fig{figitersol}. As we can see, by using the sequence of
functions approach, we can reproduce the particular exact solution to the
homogeneous equation with better and better accuracy. Finally, by solving the
inhomogeneous equation with cutoff, \eq{tKLeq}, we find that the asymptotic
behaviour of the solution $\tKL(p,k)$ is identical (up to a normalisation
factor) with the asymptotic behaviour of all the solutions $\tKLi(p,k)$ of
\eq{tKLieq}. It is thus $\tKL_h(p,k)$ to which the sequence of functions
$\tKLi(p,k)$ converges.

We now apply the above procedure to our physical case of boson-dimeron
scattering described by \eq{Keq}. Thus we first solve \eq{KLeq} with a
sufficiently large value of $\Lambda$ in order to obtain a particular solution
$\KL(p,k)$. Then we solve \eq{KLieq} with the same value of $\Lambda$ using the
sequence of inhomogeneous terms specified by \eq{Mi}. As in the simpler case
above, the asymptotic tails of $\KL(p,k)$ and all the solutions $\KLi(p,k)$ are
oscillatory and, up to a normalisation factor, identical.  This can be seen
explicitly in \fig{kh} where the $\KLi$ for $i=4$, $8$, and $10$ are plotted
after having been scaled to have the same asymptotic tail as $\KL$. From this it
is seen how the sequence of functions $\KLi$ provides successively better
approximations to the function $\KL_h$ appearing in \eq{KLstruc}. With $i$
sufficiently large, we can subtract $\KLi$ from $\KL$, in this way deducing the
desired physical $K$ matrix $K_{ph}$ with vanishing asymptotics.
\begin{figure}[t]
\centerline{\epsfxsize=8.0cm\epsfbox{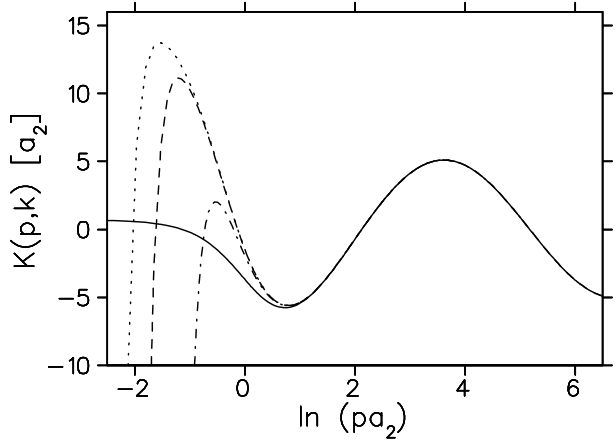}}
\vspace{1mm}

\caption{Numerical construction of the function $\KL_h(p,k)$ defined by
\eq{KLheq} and \eq{KLhsim}. The solid line is the solution $\KL(p,k)$ to the
inhomogeneous $K$ matrix equation with cutoff $\Lambda=1.6\times 10^7a_2^{-1}$
and on-shell momentum $k=0.82a_2^{-1}$. The dot-dashed, dashed, and dotted
curves correspond to sequential approximations to $\KL_h(p,k)$ with $i=4$, $8$,
and $10$, respectively.}  \fign{kh}
\end{figure}

As evident from \fig{kh}, lager and larger values of $i$ are needed in order
to approximate the value of $\KL_h(p,k)$ for progressively smaller values of the
momentum $p$. For our purposes it is sufficient to have $i$ only large enough to
give an accurate approximation to $\KL_h(p,k)$ at the on-mass-shell value of
$p$, i.e., at $p=k$. By repeating the above calculations for a range of values
of the on-mass-shell momentum, we obtain the physical $K$ matrix $K_{ph}(k,k)$
as a smooth function of $k$. The result is drawn in \fig{onshell}.
%This shows explicitly that the boson-dimeron scattering amplitude in this
%leading order EFT is determined purely by two-body input.

The above numerical procedure cannot be used directly to obtain the
boson-dimeron scattering length because our sequence of functions method to
determine $\KL_h(p,k)$ does not extend to the value of $p=0$.  However, taking
into account the smoothness of the scattering amplitude, we can determine the
scattering length by extrapolating the amplitude to zero momentum.  There is not
much sense in making that kind of extrapolation for the present scalar particle
EFT model; however, we will be able to compare the predicted value for the
scattering length with the experimental value when we complete calculations for
the case of neutron-deuteron scattering in the doublet channel.

%Armed with numerical results for solutions to the homogeneous equation we take
%numerical data for $\KL(p,k)$ with fixed value for $\Lambda $ and
%construct the following quantity:
%\be
%K_{ph}^*(p,k)=\KL(p,k)+c_1K_h^0(p,k)+c_2K_h(p,k)
%\label{prephampl}
%\end{equation}
%where $c_1$ and $c_2$ are arbitrary parameters.
%Choosing the values of these parameters appropriatelly we cancel the
%oscillating 
%bahaviour of $\KL(p,k)$ for large $p$. Obtained quantity
%\be
%K_{ph}(p,k)=\KL(p,k)+c_1^{ph}K_h^0(p,k)+c_2^{ph}K_h(p,k)
%\label{phampl}
%\end{equation}
%is the desired solution to the \eq{aeq} with vanishing asymptotic
%behaviour. This solution is plotted in FIG. \ref{onshell}. $K_{ph}(p,k)$ provides
%us with the physical amplitude $K_{ph}(k,k)$ 
%of the particle scatttering off a two-body bound state. 

\section{Summary}

Applying leading order EFT to describe the scattering of a boson off a two-boson
bound state results in an unusual inhomogeneous integral equation, \eq{aeq},
originally derived in the 1950's by Skorniakov and Ter-Martirosian \cite{ST},
whose kernel $MG_0$ is such that the operator $1-MG_0$ has no inverse. The same
equation arises in the application of leading order EFT to neutron-deuteron
scattering where its unusual properties have caused difficulties with its
numerical solution \cite{BHK_PRL}. The problem encountered in Ref.\
 \cite{BHK_PRL} is that the scattering amplitude has no limit as the cutoff,
used to solve the equation, is removed; in particular, it was found that as the
cutoff is taken to infinity, the amplitude oscillates with a constant
amplitude. It was argued that this undesirable behaviour of the amplitude should
be eliminated by the introduction of a carefully chosen three-body force into
the leading order Lagrangian \cite{BHK_PRL}.
\begin{figure}[t]
\centerline{\epsfxsize=8.0cm\epsfbox{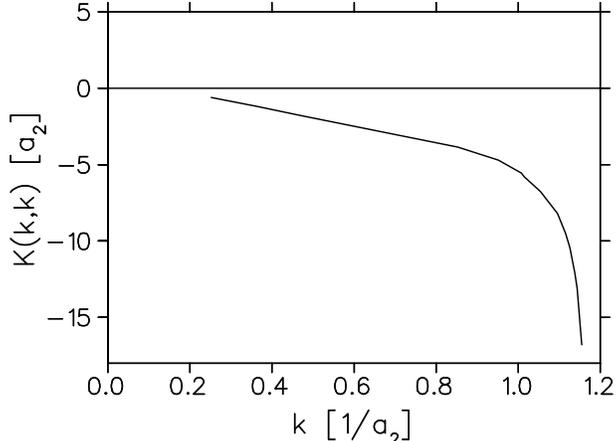}}
\caption{The on-mass-shell $K$ matrix $K_{ph}(k,k)$ corresponding to the
physical amplitude for boson-dimeron scattering. This numerical result is
based on leading order EFT with two-body input only.}
\fign{onshell}
\end{figure}

In the present paper we have solved the problem of the ultraviolet
cutoff-dependence of the scattering amplitude without the need to introduce a
three-body force. We did this by taking into account the fact that, for any
given energy, \eq{aeq} has an infinite number of solutions amongst which only
one corresponds to the physical amplitude for particle-bound state scattering.
Rather than solving \eq{aeq} directly, we chose to work with the corresponding
$K$ matrix equation, \eq{Keq}. Although \eq{Keq} still has an infinite number of
solutions, it has the advantage of singling out only those amplitudes which are
unitary.  By introducing a cutoff $\Lambda$ into \eq{Keq} and numerically
solving the resulting equation, \eq{KLeq}, we have shown that {\em every}
$\Lambda$ that is sufficiently large, results in an amplitude that satisfies the
original equation, \eq{aeq}.  Thus, the sensitivity of the scattering amplitude
to the ultraviolet cutoff $\Lambda$ is simply a reflection of the fact that one
is in this way obtaining {\em different} solutions of \eq{aeq} for every
$\Lambda$. A simple numerical analysis of our solutions to \eq{KLeq} has allowed
us to deduce the functional form of the $\Lambda$ dependence of the solution -
see \eq{Kstruc}.

Amongst the infinite number of solutions $\KL(p,k)$ described by \eq{Kstruc},
the $K$ matrix corresponding to the physical amplitude, $K_{ph}$, is
distinguished by the fact that it vanishes in the limit of infinite off-shell
momentum $p$. This asymptotic behaviour can be contrasted with the oscillatory
asymptotic behaviour of all other solutions. By introducing a sequence of
equations like \eq{KLeq} but with progressively smaller inhomogeneous terms, we
have managed to extract $K_{ph}$ numerically from our numerical solutions of
\eq{KLeq} - see \fig{onshell}. In this way we have shown explicitly that EFT in
leading order describes the scattering of a particle off a two-body bound state
in terms of two-body forces only.

%Spinless particle - two-body bound state scattering problem at leading order
%leads to the equation of Skorniakov and Ter-Martirosian together with a boundary
%condition at the origin (in configuration space) which eliminates the
%oscillating behaviour.  Hence EFT resolves quite naturally the problem of the
%choice for the arbitrary parameter which is present in the general solution to
%this equation for any values of the energy. Note that in the original approach
%by Skorniakov and Ter-Martirosian the boundary condition has to be introduced by
%hand \cite{danilov}.

%If we are working up to some higher order then to remove divergences we need to
%include contributions of
%three-body counter-terms. The resulting amplitude does not
%correspond to any two-body potential and hence Thomas's theorem does not
%apply. 

%If we keep cut-off finite then our results are reliable only up to the order we
%are 
%working with. At leading order the above discussions apply and if we are
%working 
%up to higher orders then three-body counter-terms are to be included to
%compensate  
%cut-off dependence of the scattering amplitude (up to desired order). As
%Thomas's theorem is derived without taking into account three-body forces, it
%does not apply to the
%cut-off EFT at higher orders. 

%From the above discussion one concludes that to handle the
%particle - two-body bound state scattering problem 
%one does not need to include contributions of the three-body force 
%into leading order EFT calculations as proposed in \cite{BHK_NP}.

%\section{Appendix}

\medskip
\medskip
%{\bf Appendix}
\medskip

\medskip
\medskip
\medskip

{\acknowledgements}

We would like to thank I.\ R.\ Afnan and V.\ Pascalutsa for numerous useful
discussions. This work was carried out whilst one of authors (JG) was a
recipient of an Overseas Postgraduate Research Scholarship and a Flinders
University Research Scholarship at the Flinders University of South Australia.

%\newpage 
%\begin{thebibliography}{99} 

\end{document}